\documentclass[sigconf]{acmart}

\usepackage{todonotes}
\usepackage{graphicx}
\usepackage{subfig}
\usepackage{appendix}

\makeatletter
\patchcmd{\hyper@makecurrent}{%
    \ifx\Hy@param\Hy@chapterstring
        \let\Hy@param\Hy@chapapp
    \fi
}{%
    \iftoggle{inappendix}{
        \@checkappendixparam{chapter}%
        \@checkappendixparam{section}%
        \@checkappendixparam{subsection}%
        \@checkappendixparam{subsubsection}%
        \@checkappendixparam{paragraph}%
        \@checkappendixparam{subparagraph}%
    }{}%
}{}{\errmessage{failed to patch}}

\newcommand*{\@checkappendixparam}[1]{%
    \def\@checkappendixparamtmp{#1}%
    \ifx\Hy@param\@checkappendixparamtmp
        \let\Hy@param\Hy@appendixstring
    \fi
}
\makeatletter
\newtoggle{inappendix}
\togglefalse{inappendix}

\apptocmd{\appendix}{\toggletrue{inappendix}}{}{\errmessage{failed to patch}}
\apptocmd{\subappendices}{\toggletrue{inappendix}}{}{\errmessage{failed to patch}}

\AtBeginDocument{%
  \providecommand\BibTeX{{%
    \normalfont B\kern-0.5em{\scshape i\kern-0.25em b}\kern-0.8em\TeX}}}

\setcopyright{rightsretained}
\copyrightyear{2022}
\acmYear{2022}
\acmDOI{}
\acmConference[EduCHI'22]{4th Annual Symposium on HCI Education}{April 30-May 1 2022}{New Orleans, LA, USA}
\acmBooktitle{EduCHI'22: 4th Annual Symposium on HCI Education, April 30-May 1 2022, New Orleans, LA, USA}
\acmPrice{15.00}

\begin{document}

\title[Coding IxD: Enabling Interdisciplinary Education by Sparking Reflection]{Coding IxD: Enabling Interdisciplinary Education by Sparking Reflection}

\author{Peter Sörries}
\authornote{Both authors contributed equally to this research.}
\email{peter.soerries@fu-berlin.de}
\affiliation{%
  \institution{Freie Universität Berlin}
  \city{Berlin}
  \country{Germany}
}

\author{Judith Glaser}
\email{glaser@kh-berlin.de}
\authornotemark[1]
\affiliation{%
  \institution{Weißensee School of Art and Design}
  \city{Berlin}
  \country{Germany}
}

\author{Claudia Müller-Birn}
\email{clmb@inf.fu-berlin.de}
\affiliation{%
  \institution{Freie Universität Berlin}
  \city{Berlin}
  \country{Germany}
}

\author{Thomas Ness}
\email{ness@kh-berlin.de}
\affiliation{%
  \institution{Weißensee School of Art and Design}
  \city{Berlin}
  \country{Germany}
}

\author{Carola Zwick}
\email{zwick@kh-berlin.de}
\affiliation{%
  \institution{Weißensee School of Art and Design}
  \city{Berlin}
  \country{Germany}
}

\renewcommand{\shortauthors}{Sörries and Glaser, et al.}

\begin{abstract}

Educating students from diverse disciplinary backgrounds is challenging. 
In this article, we report on our interdisciplinary course \emph{coding interaction and design} (Coding IxD), which is designed for computer science and design students alike. This course has been developed over several years by consciously deliberating on existing hurdles within the educational concept. First, we motivate the need for Coding IxD and introduce the teaching principles that helped shape the course's general structure. Our teaching principles materialize in four method-based phases derived from research through design. Each phase consists of several methods that emerged to be suitable in an interdisciplinary context. Then, based on two selected student projects, we exemplify how interdisciplinary teams can arrive at novel interactive prototypes. We conclude by reflecting on our teaching practice as essential for a meaningful learning experience.

\end{abstract}

\ccsdesc[500]{Applied computing~Education}
\ccsdesc[500]{Human-centered computing~HCI theory, concepts and models}

\keywords{Education; research through design; interaction design; reflection}


\maketitle

\section{Introduction}


The interdisciplinary teaching course \emph{coding interaction and design} (Coding IxD)\footnote{For more information on Coding IxD please visit \url{https://www.codingixd.org/}.} was developed from the observation that computer scientists and designers\footnote{The term ``design'' in the understanding of Coding IxD is related to the practical experience of the design community or discipline in the fields of product and interaction design.} 
often collaborate after their studies in various settings (e.g., start-ups), but educational courses that motivate collaborative work in computer science and design are still rare in Germany. Such collaborations have several hurdles since, for example, a shared understanding of the different perspectives on the computer science and design discipline or the responsibility of those disciplines in the design process is lacking. Although other related professions such as user experience researchers or product managers are also involved in the design process in practice, Coding IxD focuses on computer science and design. The goal of Coding IxD is to realize a rich encounter between students of computer science and design in which fundamentally different perspectives and approaches meet, exchange, and evolve.
In simplified terms, designers imagine their visions by making sketches, digital or physical prototypes to materialize their ideas. This practice poses the challenge that such an impulsive and artistic approach tries to describe something vague in its visual form; however, it remains on a largely conceptual level~\cite{petruschat2019wicked}. Restrictions concerning the realization of their designs are often not considered or seen as secondary.
By contrast, computer scientists focus mainly on the technical aspects without considering the challenges and possibilities of a design. They usually pursue an engineering approach (e.g., by adopting a user-centered design process) to make their product visions feasible for users but focus on rules that are more technical in nature~\cite{petruschat2019wicked}.
This deficiency in both disciplines requires common ``designerly ways of knowing, thinking and acting''~\cite{cross2012design}.

Hence, we address this ``deficiency'' by perceiving Coding IxD as a practical contribution to further develop the diffuse and still demarcated space between computer science and design education into interdisciplinary collaboration.
In this process, designers are supposed to deal with limitations of prototypes caused by technical means, on the one hand, and, on the other hand, computer scientists with requirements on their prototypes that emerge from the design. Both perspectives, thus, merge into functional high-fidelity (hi-fi) prototypes or, as in our course language, so-called ``neo-analogue artifacts.'' We define neo-analogue artifacts as a contextualized symbiosis of code, material, and form manifested as physical interaction between the artifact and the human that causes or creates a digital outcome. We understand neo-analogue artifacts as objects that materialize in novel interaction concepts; they serve as a functioning entity that might provoke critique of ubiquitous contexts of application~\cite{petruschat2019wicked, buchanan1992wicked} while respecting both human capabilities and vulnerabilities (e.g.,~\cite{zimmerman2009designing}).


Coding IxD takes place yearly and we invite students with varying educational backgrounds (i.e., graduate or undergraduate). In each study term, we adapt our teaching principles according to a real-world design context. The latter serves as a critical lens on current social circumstances leading to ``a form of design that pushes the cultural and aesthetic potential and role of electronic products [...] to its limits''~\cite{dunne2001design}. Accordingly, we focus on emerging design contexts, which provide students with inspiration and anticipation to spark their ideas. With these design contexts, we encourage students to incorporate a reflective lens when designing neo-analogue artifacts. Consequently, we emphasize the need for students in computer science and design to critically reflect in interdisciplinary teams on their ``values, attitudes, and ways of looking at the world''~\cite{sengers2005reflective} and how these values lead to certain decisions in the design process. The teaching principles are informed by epistemological research, such as \emph{research through design}~\cite{zimmerman2007research, frayling1993research}, \emph{reflective design}~\cite{sengers2005reflective, schon2017reflective}, and \emph{critical design}~\cite{dunne2006hertzian, dunne2001design}.

In the following, we detail how we guide our students ``through an active process of ideating, iterating, and critiquing potential solutions''~\cite{zimmerman2007research} by using five teaching principles (cf.~\autoref{sec:teach}) in four carefully coordinated method-based phases (cf.~\autoref{sec:phases}). We introduce two design contexts (cf.~\autoref{sec:contexts}) from past courses accompanied by two selected student projects (cf.~\autoref{sec:projects}) that show how visions, i.e., experimental or application-oriented, can be derived through method-based design processes leading to finely tuned prototypes. Finally, we reflect on the past years of interdisciplinary education concerning our teaching principles (cf.~\autoref{sec:reflecting}).

\section{Teaching Principles of Coding IxD}
\label{sec:teach}

We (conductors) provide our research and practical experience of theory and practice in five teaching principles of Coding IxD. We have steadily improved these phases over the last years. The theoretical background of our teaching principles provides enough flexibility for our method-based phases (cf.~\autoref{sec:phases}).

We acknowledge that working with students of different disciplines and levels requires a certain sensitivity. Hence, we designed Coding IxD to facilitate ``critical reflection''~\cite{sengers2005reflective} from students and us throughout each teaching course. 
We perceive reflection as a catalyst for the design of novel prototypes by extending students’ knowledge through our method-based phases related to design. Furthermore, we consider the practice of design as a symbiosis of the culture of inquiry and action in interaction design~\cite{nelson2014design}. This process of reflection is not about solving a specific problem but more about formulating problems in terms of intentional actions~\cite{nelson2014design} leading to delightful user experiences. In this sense, students in computer science and design carry out Coding IxD as a combination of research and practice. Through the latter, we engage students to reflect on the feasibility of their prototypes through an iterative guided design process. We derived our five teaching principles from the criteria of research through design~\cite{zimmerman2007research}:\\

\begin{enumerate}
    \item \emph{Assuring Rigor.} Students apply appropriate design methods and techniques and are taught rationales for their selection. We orchestrate these methods into four method-based phases (cf.~\autoref{sec:phases}) that enable a design process from a vision to an implementation in a flexible manner.
    \item \emph{Providing Relevance.} Students articulate the preferred state of their prototypes (i.e., the design approach of a neo-analogue artifact). The goal is to reason why the research and design community, potential users, or other stakeholders should consider their state to be preferred.
    \item \emph{Ensuring Extensibility.} Students document their knowledge learned by explaining their neo-analogue artifacts to a technical and non-technical audience. The main goal, besides making design processes and artifacts accessible and reproducible, is to embark on public awareness (e.g., through interactive exhibitions).
    \item \emph{Explicating Invention.} Students take a ``glimpse'' into the future by describing delightful user experiences and materializing them in neo-analogue artifacts. The contribution aims to advance the current state of the art in the research and design community. By articulating their invention, students must detail how their design concepts result in a significant advancement.
    \item \emph{Enabling Value Work.} Students reflect on existing needs and underlying values as an essential part of the design process. Through reflection, students develop a critical stance during the design process and on potential users or other stakeholder values of a design context.
\end{enumerate}

\section{Phases of Coding IxD}
\label{sec:phases}

The teaching principle \emph{Assuring Rigor} consists of four method-based phases. These phases frame each teaching term methodically --- from the introduction of a design context and the ideation of novel prototypes through accompanying workshops, to the iterative process of concept development, to the final neo-analogue artifact. 
Each phase contains various methods that are flexible in their combination during each teaching course. Such flexibility is necessary due to the varying skills and capabilities of the students.

Phase~1, \emph{Enabling the Interdisciplinary Work}, aims to set the thematic frame for the course and to form the interdisciplinary teams. In phase~2, \emph{Establishing the Design Process}, we focus on creating a common working process and exploring the design context. The central concern of phase~3, \emph{Exploring the Concept}, is the iterative and exploratory elaboration of various ideas and testing of different design approaches. The phases conclude with phase~4, \emph{Enabling the Discourse}, with the realization of the most convincing concept by implementing the neo-analogue artifact and preparing a public presentation. In reality, of course, these phases might not be clearly separated from each other but instead overlap.

\begin{figure*}[t]
  \includegraphics[width=\textwidth]{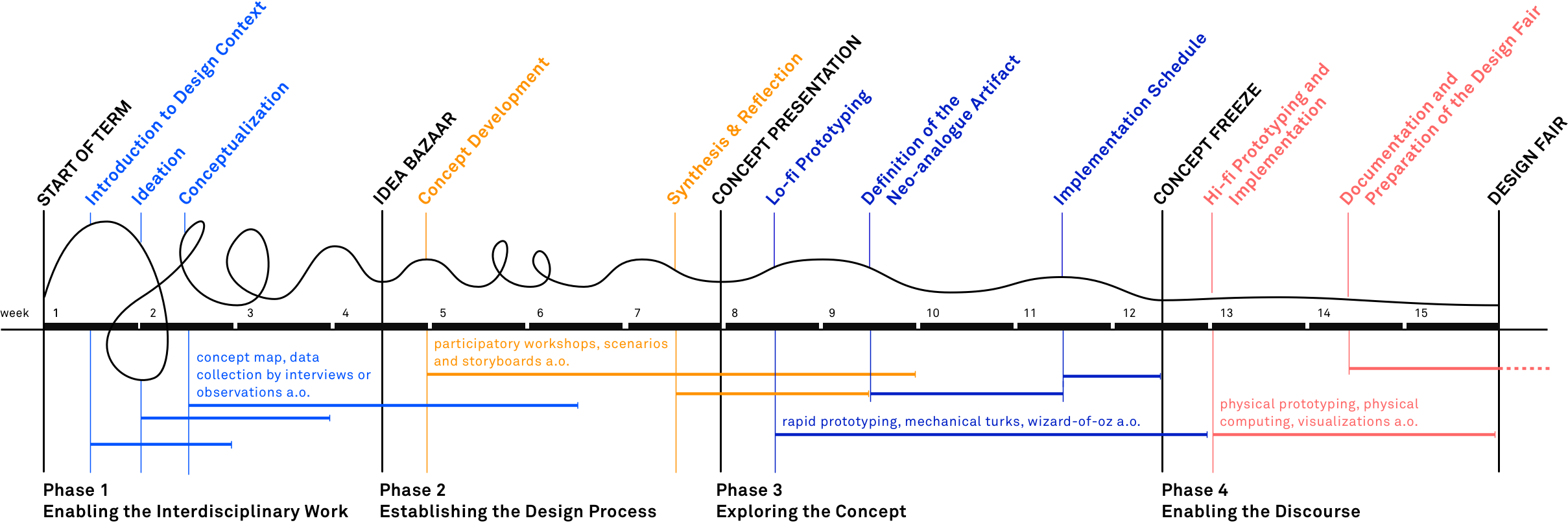}
  \caption{Visual overview of the four method-based phases with the milestones incorporated.}
  \Description[Visual overview of the four method-based phases]{Visual overview of the four method-based phases with the milestones incorporated.}
  \label{fig:phases}
\end{figure*}

We define the milestones \emph{idea bazaar}, \emph{concept presentation}, \emph{concept freeze}, and \emph{design fair} as the goal for each phase. In the following, we describe the four method-based phases incorporating the milestones in more detail (cf.~\autoref{fig:phases}).


\paragraph{Phase 1: Enabling the Interdisciplinary Work}
Building a shared understanding among students and creating a sense of commonality is a vital prerequisite for Coding IxD. Thus, the major goal of the first phase is to get both disciplines acquainted with the perspective of the other. This phase spans at least four weeks and starts with an introduction of the design context.
We dedicate a huge amount of time preparing each semester in advance, i.e., familiarizing ourselves (conductors) with the topic we have chosen for the teaching term by, for example, reviewing the literature, observing relevant social and political discourses, and inviting suitable guest lectures that introduce different perspectives on a topic. This preparation is reflected in the introduction of each course.
After this introduction, we invite students to conceptualize their knowledge, interests, and views of the design context given.
Here, we encourage students to reflect and not limit themselves by using a flexible set of methods to assess broader views of the design context, for example, by creating large-scale concept maps in small mixed groups consisting of at least one student in computer science and design. Students are encouraged to identify potential stakeholders or existing challenges in relation to the design context. They enrich their concept maps by semi-structured interviews, surveys, observations, or existing research. These concept maps are used as intermediaries to facilitate the communication between the different disciplinary backgrounds.
This first phase ends with a talk given by each student individually and discussion of the specific area of interest chosen from these concept maps within the so-called \emph{idea bazaar}. We developed the concept of an \emph{idea bazaar} since we observed that students appreciate the ``freedom'' of choosing their topic, even though we apply restrictions (e.g., size, skill set) on the composition of each interdisciplinary team. The \emph{idea bazaar} results in interdisciplinary teams that share a common interest.

\paragraph{Phase 2: Establishing the Design Process}
In the second phase, we establish a shared working process and further conceptualize the design context, i.e., area of interest. The teams focus on a reflected (re)formulation of their chosen area of interest and related design tasks. Through weekly team spotlight presentations followed by discussion, we ensure that problems are clearly defined according to their topic, rather than thinking of ``products.'' We stimulate this ideation process by a further conceptualization through workshops or interviews with potential stakeholders, which helps the students to sharpen their vision. Consequently, we support students to illustrate their envisioned analogue and digital features of the user experience. Students then synthesize and reflect on the various perspectives of their concept by using personas, scenarios, or user journeys and, finally, they prepare the \emph{concept presentation}. Experts from research and design are invited to provide feedback on the design concepts.

\paragraph{Phase 3: Exploring the Concept}
In the third phase, students explore their different design concepts within a highly iterative process. The teams then specify their final prototype or neo-analogue artifact, i.e., the minimal viable product (MVP) that exhibits the major analogue and digital features in the later \emph{concept freeze}. We accompany this process with intensive supervision and regular group presentations to ensure that the teams reflect critically on the quality of their design concepts. The students examine the possible analogue and digital capabilities of the envisioned neo-analogue artifact during low-fidelity (lo-fi) prototyping. The MVP is then defined based on a careful selection process. This MVP will enable people to experience the unique capabilities and user experience of the product envisioned. Thus, the students specify the implementation schedule of the MVP and the materials needed for the design fair, finally captured in the \emph{concept freeze}.

\begin{figure*}[t]
  \centering
  \includegraphics[width=0.495\textwidth]{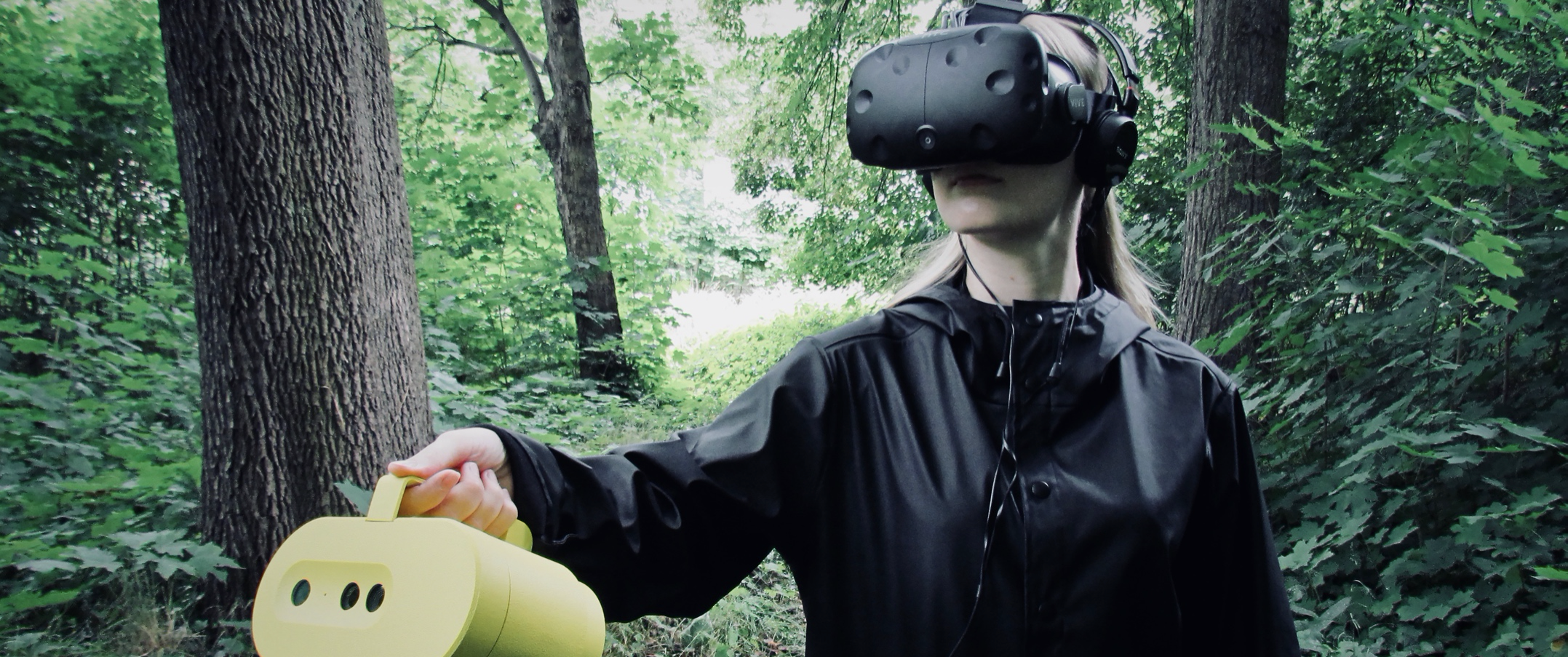}
  \hfill
  \includegraphics[width=0.495\textwidth]{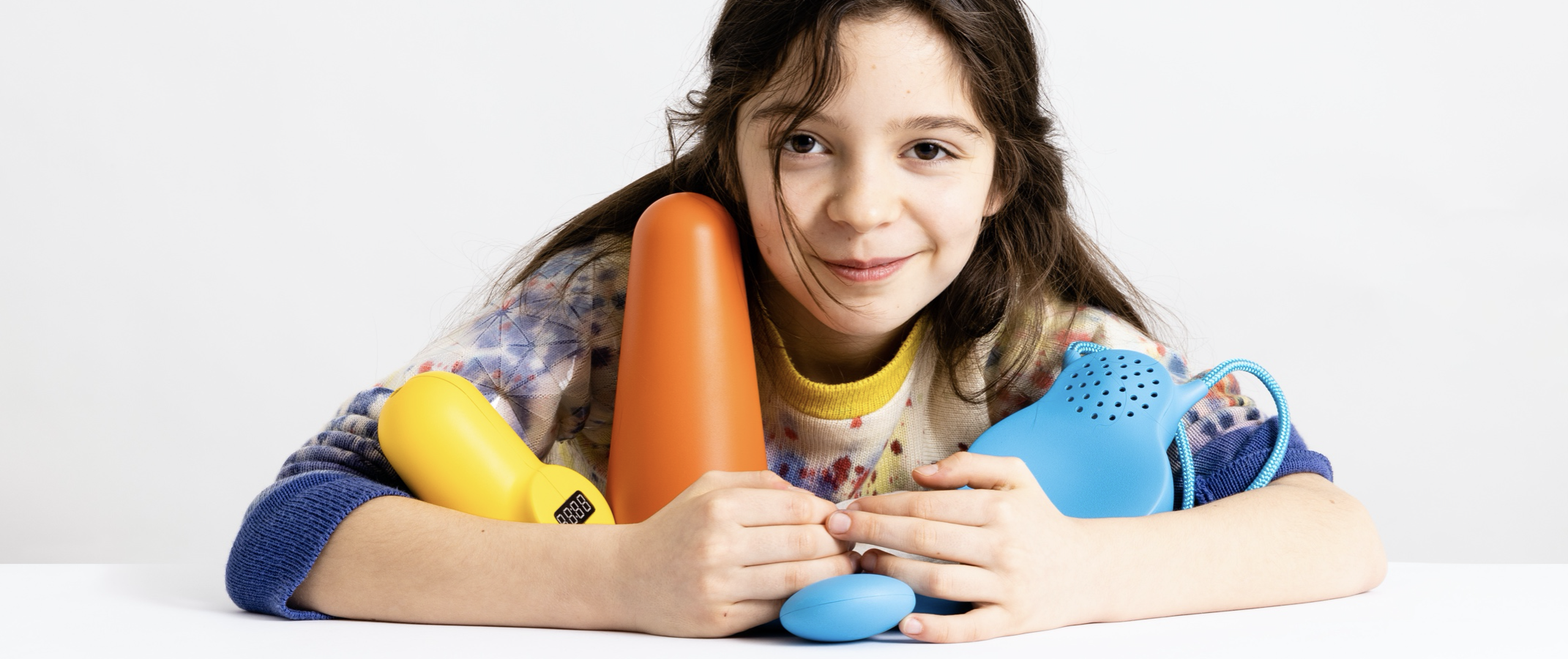}
  \caption{Selected student projects: Aux Synesthesia (left) explores physical and virtual realties through the acoustic resonances of objects channeled in novel neo-analogue user experience. Aktiv Labor (right) educates schoolchildren on traffic safety through three neo-analogue artifacts that address and train a specific cognitive skill.}
  \Description{Selected student projects: Aux Synesthesia explores physical and virtual realties through the acoustic resonances of objects channeled in novel neo-analogue user experience. Aktiv Labor educates schoolchildren on traffic safety through three neo-analogue artifacts that address and train a specific cognitive skill.}
  \label{fig:projects}
\end{figure*}

\paragraph{Phase 4: Enabling the Discourse}
In terms of the neo-analogue artifact, students continue to sharpen their design concept for the \emph{design fair} during prototyping. Given the design context, we encourage students to reflect on the potential impact of their neo-analogue artifact in light of existing political or sociocultural debates. Based on this, decisions are made regarding what technology is used and how to realize this in the MVP; by considering the teaching principle \emph{Ensuring Extensibility}, much of the effort is on the documentation and reproducibility of the neo-analogue artifact. The ongoing design process consisting of, for example, observations and interviews from phase~1, prototypical experiments, and insights from course discussions will be recorded in a research diary. This research diary of each team is intended to support reflective thinking about design decisions and activities by documenting materials such as data collection from interviews, prototype videos, and formal design studies. In addition, all code will be provided with an open-source license on GitHub. With this in mind, we prepare the teams for the \textit{design fair}, where we aim to provoke a lively discourse about the results both internally and with the broader public. The \textit{design fair} is realized as an interactive exhibition format that allows each team to showcase their embodied designs to the public. Visitors can explore and experience the functional artifacts and discuss them with the teams.

\section{Design Contexts}
\label{sec:contexts}

We have explored several design contexts over the past years: using data from cultural institutions, exploring the impact and capabilities of technologies such as the internet of things, mixed reality, machine learning algorithms, or exploring the digital nature of the urban space. We reflected on existing design potentials concerning the current political or sociocultural debates in every design context.
In the following, we introduce two representative design contexts: one focusing on exploring mixed realities and the other on urban spaces. We selected one project for each design context to show how a design process evolves and deviates from an obvious or straightforward solution enabling new and even unexpected perspectives according to our teaching principles and methodological underpinning.

\paragraph{Design Context 1: Exploring Mixed Realities}

Mixed realities including augmented reality~(AR) and virtual reality~(VR) have become very popular in recent years. These innovations found their commercial use in a non-technical audience, especially in the entertainment industry. In the 1980s, VR became known as the ``vision of a new kind of computer experience''~\cite{cockayne2003virtual} of three-dimensional environments, leading via the first computer-aided design courses for engineers in the 1990s to the fully virtual world we know today. Nowadays, for example, people can be fully immersed in artificial realities that address a variety of human sensory modalities or even weave multiple realities together and perform interactions in real-time. But would it not be more satisfying to design holistic experiences as the technological capabilities of AR/VR can create multilayered and customized multi-sensory experiences; by addressing the nature of augmentation through human actions and perceptions? As the quality of embodied interaction principles, the presence or absence of gravity, and other tacit perceptual phenomena inspire thinking about interactive experiences for hybrid realities through novel artifacts.

\paragraph{Design Context 2: Neo-analogue Products for Urban Spaces}

Increasing digitization is impacting our cities. Networks of mobile devices, sensors, actuators, and intelligent algorithms that collect and analyze urban data in real-time are at the heart of such ``smart cities''~\cite{freeman2019smart}. However, purely technologically focused approaches usually fail because the specific needs of the city and its inhabitants are neglected. Designing for smart cities is more than implementing urban computing technologies~\cite{paulos2004familiar} and developing more sophisticated data analytics. Cities function as finely tuned psycho-geographic entities that influence the perceptions, psychological experiences, and behaviors of their inhabitants~\cite{van2010cities}. Urban informatics, therefore, focuses on the social and human impact of technology~\cite{freeman2019smart} by fostering the new culture of neo-analogy through the development of smart city strategies that can incorporate meaningful intersections between people, places, stories, purposes, and technologies. These meaningful intersections can only be achieved if the identity of a city is taken into account. This design context intends to uncover, explore, and shape these intersections in urban spaces.

\begin{figure*}[ht]
    \centering
    \begin{minipage}[t]{26.75em}
        \centering
        \includegraphics[width=0.325\textwidth]{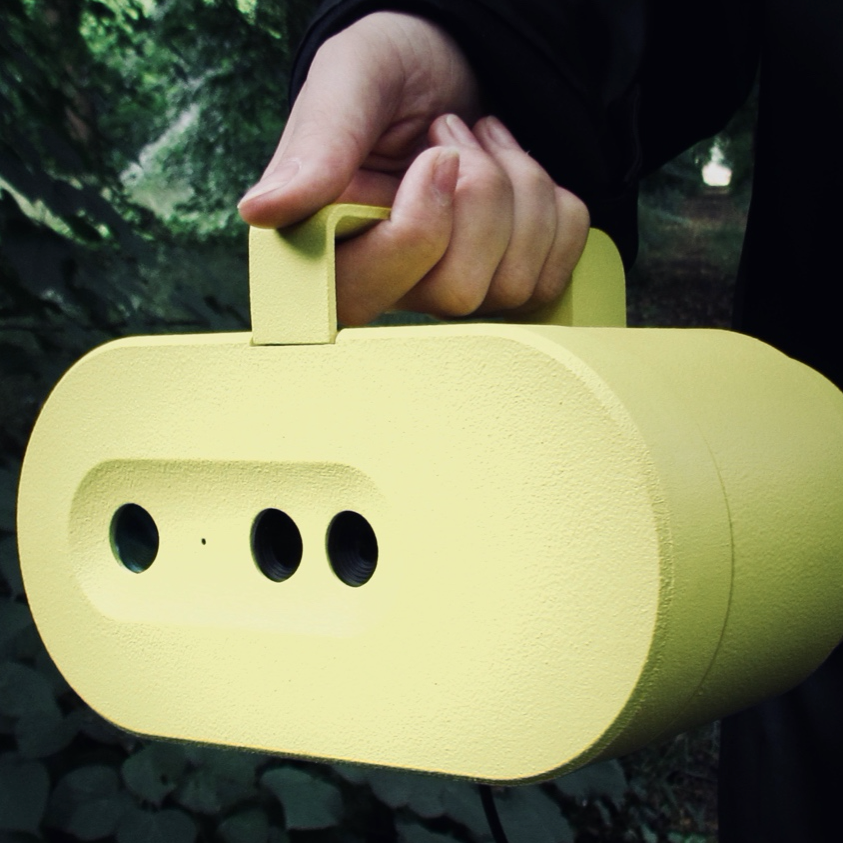}
        \hfill
        \includegraphics[width=0.325\textwidth]{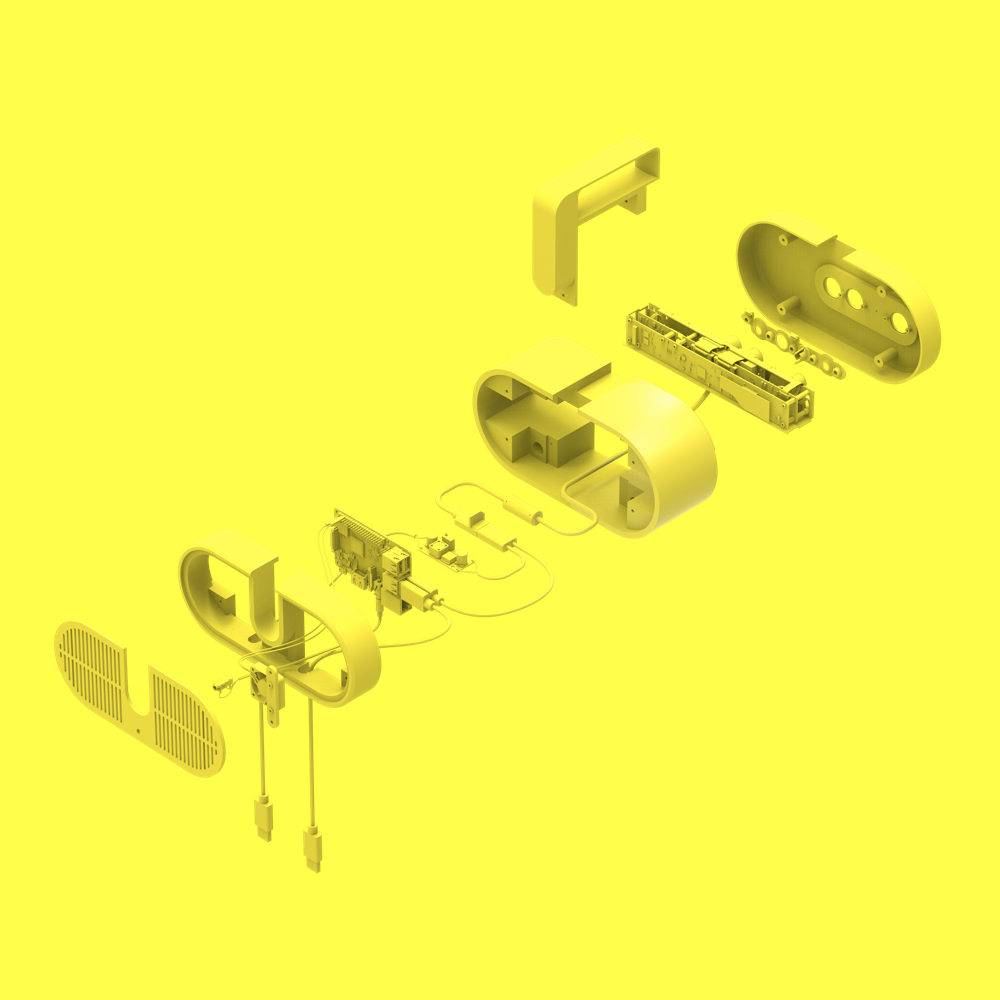}
        \hfill
        \includegraphics[width=0.325\textwidth]{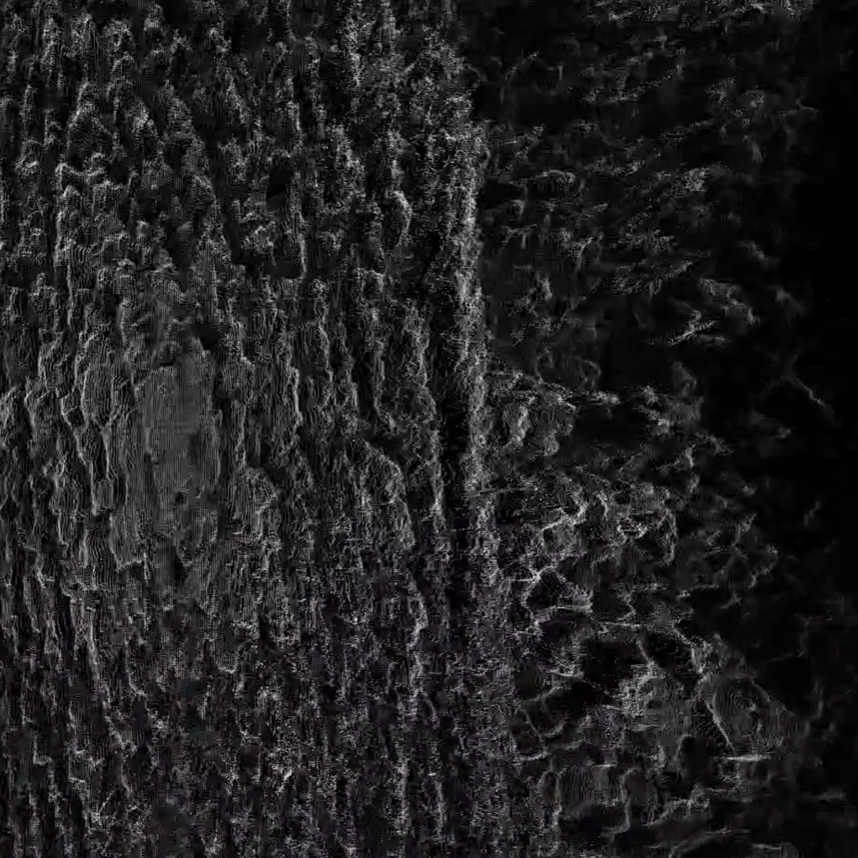}
        \caption{Portable artifact (left) for spatial exploration maps the physical environment and processes the data mapped. The disassembled portable artifact (middle) consists of a sensor for data acquisition and a single-board computer for data processing. Abstracted spatial reality (right), visible through a head-mounted display, is based on collected and processed spatial data.}
        \Description{Portable artifact for spatial exploration maps the physical environment and processes this data mapped. The disassembled portable artifact (middle) consists of a sensor for data acquisition and a single-board computer for data processing. Abstracted spatial reality (right), visible through a head-mounted display, is based on collected and processed spatial data.}
        \label{fig:auxsynesthsia}
    \end{minipage}
    \hfill
    \begin{minipage}[t]{26.75em}
        \includegraphics[width=0.325\textwidth]{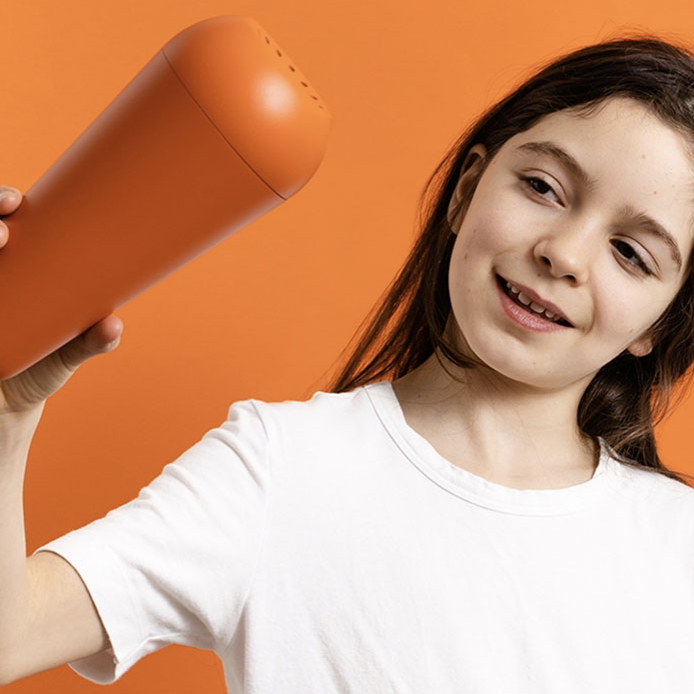}
        \hfill
        \includegraphics[width=0.325\textwidth]{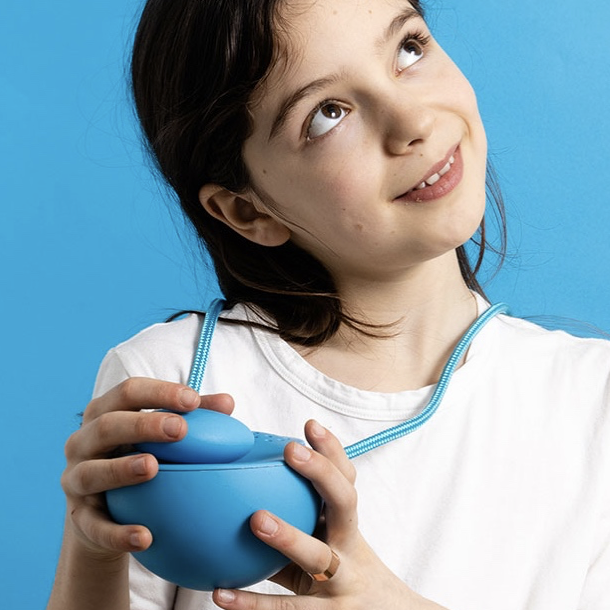}
        \hfill
        \includegraphics[width=0.325\textwidth]{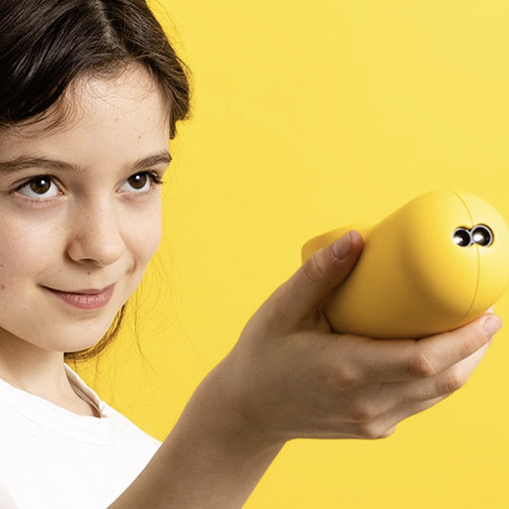}
        \caption{\emph{Hörbius} (left) enhances the directional hearing skills of schoolchildren. \emph{Orienta} (middle) finds the safe way and trains navigation and orientation skills. \emph{Distanzo} (right) estimates distance and conveys the speed of traffic vehicles.}
        \Description{Hörbius enhances the directional hearing skills of schoolchildren. Orienta finds the safe way and trains navigation and orientation skills. Distanzo estimates distance perception and conveys the speed of traffic vehicles.}
        \label{fig:aktivlabor}
    \end{minipage}
\end{figure*}

\section{Selected Student Projects}
\label{sec:projects}

In light of the design contexts, we present two student projects (cf.~\autoref{fig:projects}) to validate our teaching principles and method-based phases of Coding IxD in~\autoref{fig:refstud}.

First, the Aux Synesthesia project originated in the design context \emph{Exploring Mixed Realities}. In this case, the students were inspired by the acoustic resonance of objects and their materiality (i.e., texture and surface) addressing human senses in unforeseen ways.
Here, the project team applied the method-based phases of Coding IxD to their purposes in more exploratory and artistic ways to channel their ideas (e.g., by performing certain acoustic tests in different environments and comparing the results in a proof of concept to map design features into lo-fi prototypes). This methodologically adapted approach affected immensely the later user experience and formal aesthetics of the physical and visual artifacts.

The Aktiv Labor project, from the design context \emph{Neo-analogue Products for Urban Space}, investigated how schoolchildren can be prepared and educated responsibly for traffic safety in urban areas.
The Aktiv Labor team used a more user-centered approach by conducting several participatory workshops in elementary school classrooms. Initially, the workshops were used to explore the schoolchildren's individual positive and negative experiences on their way to school. These were followed by playful approaches to assess the cognitive abilities of schoolchildren. As the project progressed, the purpose of the workshops shifted to evaluating and refining elaborated concepts using lo-fi prototypes and their design features (technological and formal aspects) finally resulting in three neo-analogue artifacts.

\subsection{Aux Synesthesia}

Humans are visual beings and use their senses without restriction. Humans have become accustomed to this state and unconsciously react to a number of impressions and overlook subtle changes. Bats, for example, their emitted echoes are reflected by objects and sent back to them. The emitted echoes allow bats to estimate distances or even locate the position and shape of specific objects. But how can sensory perception be reinterpreted and explored by humans shaping novel spatial experiences?

The Aux Synesthesia team, consisting of two computer science students and one design student, is taking advantage of this vision to cause changes in the cognitive abilities humans take for granted. The project creates an interface between real and virtual environments, in which acoustic signals traverse space and are absorbed, reflected or altered in their quality. In the course of analogue experiments conducted in advance, it turned out that objects have individual acoustic characteristics caused by their materiality, i.e., texture and surface. Each object has its own individual ``echo'' and, thus, distinguishes from other objects. These observations are translated as auditory dimensions into an extraordinary virtual space based on the idea of reducing and reinterpreting the physical reality, changing our perception how to ``see'' things in a tangible experience.

Based on the synthesis of a physical prototype, the real space is read via a portable artifact (cf.~\autoref{fig:auxsynesthsia}, left and middle). Comparable to a Geiger counter, the environment is mapped through an incorporated Microsoft Kinect sensor in real-time and processed through a RaspberryPi single-board computer.
Approaching objects are detected and interpreted as digital stereo signals. These signals or frequency levels adapt to the current conditions of objects (e.g., distance, size, or materiality). This acoustic scenario is translated into an abstracted reality (cf.~\autoref{fig:auxsynesthsia}, right) created with the game engine Unity, which allows the user of a head-mounted display to immerse themselves in the hybrid and auditive reality designed.

\subsection{Aktiv Labor}

Cities and their transportation landscape are constantly evolving. This especially poses a high risk to schoolchildren and their daily way to school through narrow or heavily frequented streets. Although traffic education is an essential part of the curriculum for schoolchildren in Germany, the teaching materials are mostly fit for purpose. Traffic safety, especially from the perspective of schoolchildren, can be taught more sustainably through playful approaches.

The team Aktiv Labor took this approach of playful learning and combined modern technologies for interactive traffic education in the school context. The team, consisting of two computer science students and three designers, developed an interactive toolkit that supports teachers to convey skills for traffic safety to schoolchildren. Aktiv Labor allows intuitive learning on an individual basis or in small teams within the safe space of a class room.

The three tools (neo-analogue artifacts) support schoolchildren’s sensory perception to facilitate various cognitive skills (e.g., hearing, seeing, and estimating distances). The three tools were refined during several design phases, including ``shadowing'' activities (e.g., co-creation workshops) to gain insights on how schoolchildren take their way to school. Not only the shape and the intuitive use of each tool play a decisive role but also the technological interface represented by them. Thus, sensors for auditory noise detection and emission or distance measurement were implemented through Arduino microcontrollers to make them functionally convincing.
\emph{Hörbius} (cf.~\autoref{fig:aktivlabor}, left) teaches directional listening skills by simulating familiar street sounds. The latter helps schoolchildren learn the type of sounds, such as those of a car, assessing potential hazards in urban spaces.
\emph{Orienta} (cf.~\autoref{fig:aktivlabor}, middle) trains navigation and orientation skills by helping schoolchildren practice routes, such as the way to school, in a comprehensible way. So-called digital markers are distributed that communicate with the artifact to support the schoolchildren in finding a safe way.
\emph{Distanzo} (cf.~\autoref{fig:aktivlabor}, right) shapes distance perception and conveys the feeling of speed and braking distances of vehicles in public traffic. The device makes it possible to estimate and check distances immediately.

\subsection{Interdisciplinary Practices within Student Projects}
\label{fig:refstud}

Although the outcomes of both student projects and their vision (i.e., experimental and application-oriented) differed obviously in their design processes, the projects nevertheless are unified by the method-based phases, the accuracy of the elaboration, and the high motivation of both teams. In the following, we describe how these projects evolved through vision and reflect on each design concept of the two projects.\footnote{For a more comprehensive assessment of the two design processes, we refer to the visual comparison as shown in~\autoref{sec:visualComparison} (Figure~5).}

The Aux Synesthesia team explored their idea to reinterpret human sensory perception in a widely explorative approach. The team was able to create a consistent experience evaluated through several experiments. These experiments (e.g., through acoustic or VR experiments) during phase~1 and~2 enabled the critical reflection and steadily reconsideration of each prototype developed and its functional performance in phase~3. The team blended together the knowledge and insights gathered during the course, from both a technical and design perspective, to create a coherent and functioning interaction concept in phase~4.
As mentioned in the vision of Aux Synesthesia, human sensory perception goes beyond the general understanding of how humans can use their senses. The initial definition of human sensory perception was therefore no longer acceptable and had to be thought of as a new integrated whole in an explorative ideation process. Here, an idea emerges from lived and observed experience, which is embodied, set in motion, and shaped through iterative prototyping (cf.~\cite{petruschat2019wicked}). Exploration and integration of experiences become a method-based design process, driven by the students' vision and the experiences gathered during the course.

The team of Aktiv Labor formed under the umbrella of a joint interest in improving urban space for schoolchildren. By consistently targeting this group across all four phases and comprehensively exploring the context and potential stakeholders involved, a thoughtful and desirable solution was developed. 
The user studies in the form of workshops in phase~1 led to a grounded understanding of how schoolchildren perceive urban traffic and safety. Here, workshops served to explore the field of interest in design and computer science concerning the design context.
The results of the co-creation workshops in a classroom setting became apparent that enabled the team to refine the objective but also to envision the relevance of their work as ``[...] interventions that utilize sophisticated, interdependent claims about fit entail complex reasoning about means and ends''~\cite[p. 5]{harrison2007}. In subsequent co-creation workshops, the Aktiv Labor team reviewed and iterated interaction concepts and prototypes. This approach significantly shaped the team's understanding of a classroom as a highly dynamic space involving multiple goals (of schoolchildren and teachers) in which learning is more than a ``transfer of information.''

Comparing both projects, Aux Synesthesia successfully utilized the prototypes for exploration and alignment of a novel user experience. Whereas Aktiv Labor primarily used several workshops as a ``vehicle for discussion'' with the user group and stakeholders involved.
The ``neo-analogue'' developed in both projects represents artifacts that bear particular witness to the students' understanding of a given design context, thus, establishing their validity beyond mere ``novelty.''

\section{Reflecting on our Teaching Practice}
\label{sec:reflecting}

In Coding IxD, we challenge students by getting to know the other disciplines and cultures of knowledge, exploring the strengths and limits of one’s discipline (i.e., articulating and arguing their creative position), and discussing the limits of democratic processes (i.e., who has the final say in a design: all or one).
%
%
%

We observe disciplinary differences, for example, how students perceive their ways of working and adapt their skills and expertise. Therefore, in \emph{Assuring Rigor}, we establish guidance on a methodological basis to encourage students to think ``outside of the box.''
We reinforce critical reflection of the design process as a vital prerequisite for applying appropriate design methods. Accordingly, differences can only be overcome if students’ needs are critically reflected and negotiated, individually, in the whole course, and in their teams.
\emph{Providing Relevance} frames the design context (e.g., defines first challenges and considers multiple perspectives on a specific situation, event, or technology).
The two projects selected illustrate how mutual interdisciplinary exchange inspires and motivates novel neo-analogue artifacts. We experience exciting projects and design processes inventing new possibilities in each course.
Such a valuable working experience enriches the design process. We perceive sound evidence in this experience as both teams (Aux Synesthesia and Aktiv Labor) worked on projects far beyond the course. Both teams were motivated by the relevance of their visions and, thus, \emph{Ensuring Extensibility} by preparing their projects for a wider audience.
For example, the team Aktiv Labor has received a user experience award nomination, and the team Aux Synesthesia presented its invention to an international audience on well-known design blogs and at exhibitions.
However, in each course, we have to push every team to its limits to believe in their vision and encourage them to keep on track. Designing neo-analogue artifacts for a specific design context can be exhausting, and frustrate teams to iterate them steadily. One reason can be that lo-fi prototypes won't meet the requirements of a target audience or are not feasible due to the technological complexity in the short period of the course. We see it as our task to motivate teams to consider other perspectives, for example, to realize a certain technological feature that meets a suitable design.
This underlines, in addition, our idea of \emph{Explicating Invention} as students' are encouraged to take a ``glimpse'' into a future through Coding IxD to advance the current state of the art.
In addition, we integrate a negotiation space to support the anticipation of students' values and expertise in a particular design context; we refer to this level of reflection for \emph{Enabling Value Work}.

In the beginning of each course, students from both disciplines are completely unfamiliar with the other discipline’s expertise or respective ways of working. Despite all the benefits described, we are recurrently witnessing perplexity and frustration within individual teams.
Common reasons include biases and automatic role assignments, subliminal values, or leaky shared mental modalities and/or visions. Students or the interdisciplinary teams that lack these fundamentals can usually not adjust to collaborative work leading to separately acting forces, inconsistencies, and immature results.
For example, we repeatedly observed computer science students with a ``wait-and-see'' attitude. Only after the design students have worked out the design concept do the computer science students actively participate by investigating the design concept for its feasibility. We assisted teams where design concepts on the side of computer science students were discarded as ``not feasible'' or ``too complex'' without discussing basic design approaches in-depth with the design students. Therefore, in recent years, we increased the inclusion of computer science students in developing design concepts (e.g., through exercises such as storyboard creation and user journeys created by all team members). We also trained the design students to position their design decisions and to argue, for example, why a particular concept is worth the ``extra effort'' and is feasible (e.g., through appropriate prototyping sessions). Likewise, computer science students argued to be perceived as ``implementers,'' i.e., merely executing the design concepts of the design students without being involved in the design decisions. The ongoing challenge for us (conductors) is to ensure an appropriate balance, appreciation, and understanding between both disciplines, to unite both forces, and methodically integrate them into Coding IxD.
Fortunately, these problems affect only a few teams. Most computer science students told us how inspired they were by design in practice. Concerning an iterative design process and how it will enrich their working expertise, they acknowledged learning about functionality, materiality, texture, quality, or sustainable production techniques of artifacts from the design perspective.
Design students mentioned learning more about the methodological approach and skills in computer science, such as what it means to reach the limits of the feasibility of technologies.

This reinforces Coding IxD and its value of interdisciplinary collaboration to coeducate students with a meaningful and respectful attitude and an open mindset towards other disciplines.
%
%

\section{Conclusion}
\label{sec:conclusion}

In this paper, we share our understanding of interdisciplinary teaching, in which students of computer science and product design do not perceive each other like pairs of opposites but rather as mutual enrichment. In more than five years of experience teaching Coding IxD, interdisciplinarity is not a matter of course and is still a challenge for us in its realization.
Coding IxD requires a diverse place where negotiation, discussion, and error within disciplines are allowed and can take place. Compared to monodisciplinary environments, establishing such a place costs students and us additional effort. However, despite occasional exceptions, the quality of the outcomes and the individual learning experiences at the end of each course convey the added value of interdisciplinary work.
%
%
%
%
We suggest that more practical and methodical exercises in values of work are needed to further improve interdisciplinary collaboration and, thus, the results in a desirable future.  
In summary, ``design is a statement''~\cite{petruschat2019wicked} that is the result of an interdisciplinary design process, as taught and advocated in Coding IxD, to unveil novel design visions.

\section{Acknowledgments}

We thank the reviewers for their insightful comments. The interdisciplinary course Coding IxD is conducted in cooperation with the Human-Centered Computing Research Group (Freie Universität Berlin, Germany) and the Embodied Interaction Group (Weißensee School of Art and Design Berlin, Germany). In addition, Coding IxD is part of the Matters of Activity of the Cluster of Excellence (Humboldt-Universität zu Berlin, Germany). Furthermore, we would like to thank the students of the team Aux Synesthesia (Jan Batelka and Thushan Satkunanathan) and the students of the team Aktiv Labor (Katharina Bellinger, Jacob Sasse, Mattis Obermann, Tim Feige, and Marisa Nest) for providing their projects for this work.

\bibliographystyle{ACM-Reference-Format}
\bibliography{references}

\onecolumn

\appendixpage
\appendix

\section{Visual Comparison of the Design Processes}
\label{sec:visualComparison}

\begin{figure*}[h]
  \includegraphics[width=\textwidth]{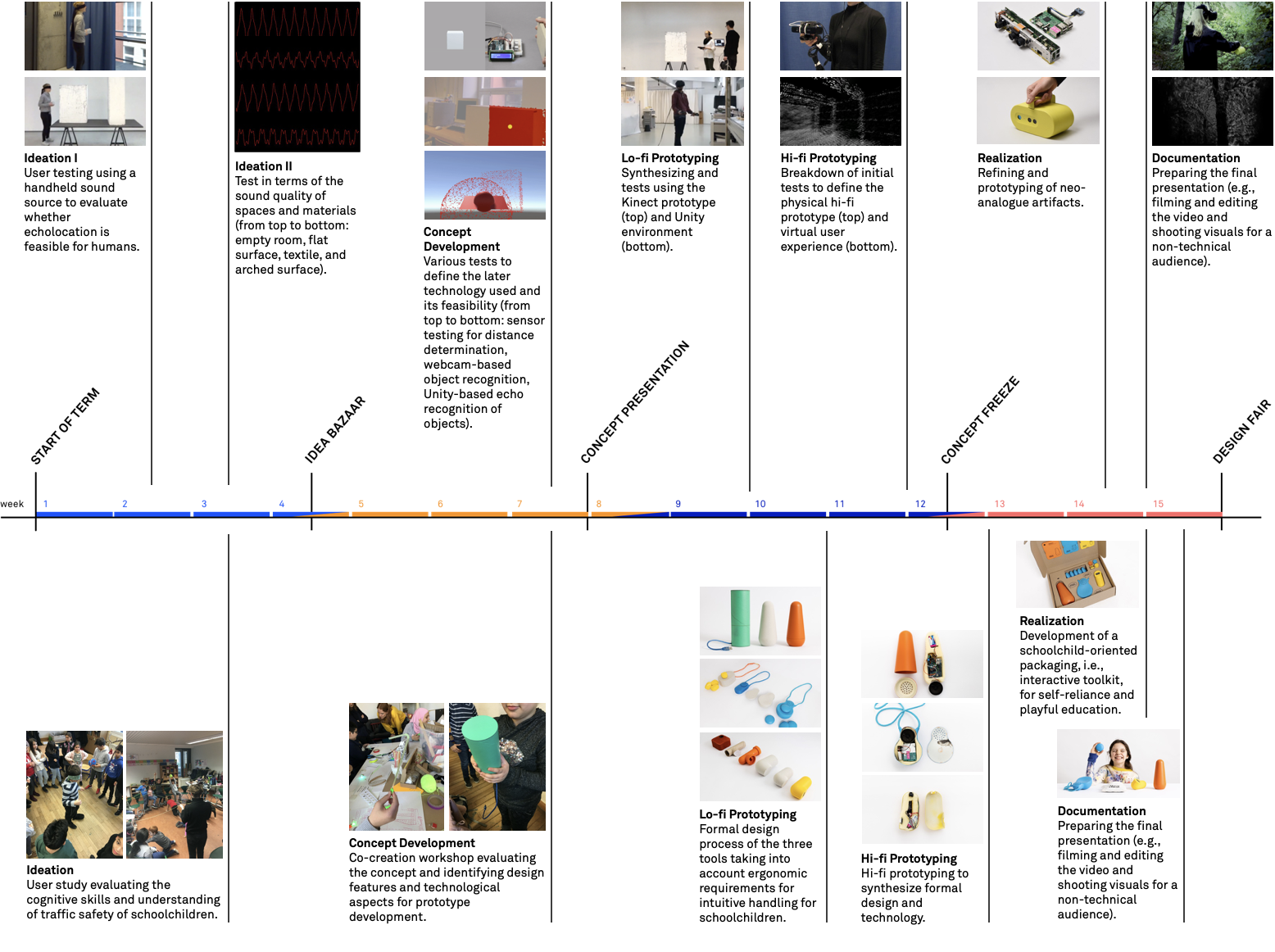}
  \caption{Visual comparison of the two student projects, showing at their core the four method-based phases and the milestones achieved through the flexible but purposeful use of each method in each phase.}
  \Description[Comparison of the design process]{Visual comparison of the two student projects, showing at their core the four method-based phases and the milestones achieved through the flexible but purposeful use of each method in each phase.}
  \label{fig:comparison}
\end{figure*}

\end{document}